\pdfoutput=1

\documentclass[conference]{IEEEtran}
\usepackage{mathtools}
\usepackage[T1]{fontenc} 
\usepackage[utf8]{inputenc} 
\usepackage{graphicx}
\usepackage{color}
\usepackage{enumerate}
\usepackage{hyperref}
\let\labelindent\relax
\let\subparagraph\relax
\usepackage{enumitem}

\usepackage[compact]{titlesec}

\begin{document}
\font\myfont=cmr12 at 22pt
\title{{\myfont Continuous Melody Generation via Disentangled Short-Term Representations and Structural Conditions}}

\author{

\Large{\shortstack{Ke Chen$^{1,2}$ \qquad Gus Xia$^2$  \qquad Shlomo Dubnov$^1$}} 

\\
\\
\large
$^{1}$CREL, UC San Diego, California, USA \\
\large
$^2$Music X Lab, NYU Shanghai, Shanghai, China \\
$^1$\{knutchen, sdubnov\}@ucsd.edu,$^2$gxia@nyu.edu
}

\maketitle

\begin{abstract}
Automatic music generation is an interdisciplinary research topic that combines computational creativity and semantic analysis of music to create automatic machine improvisations. An important property of such a system is allowing the user to specify conditions and desired properties of the generated music. In this paper we designed a model for composing melodies given a user specified symbolic scenario combined with a previous music context. We add manual labeled vectors denoting external music quality in terms of chord function that provides a low dimensional representation of the harmonic tension and resolution. Our model is capable of generating long melodies by regarding 8-beat note sequences as basic units, and shares consistent rhythm pattern structure with another specific song. The model contains two stages and requires separate training where the first stage adopts a Conditional Variational Autoencoder (C-VAE) to build a bijection between note sequences and their latent representations, and the second stage adopts long short-term memory networks (LSTM) with structural conditions to continue writing future melodies. We further exploit the disentanglement technique via C-VAE to allow melody generation based on pitch contour information separately from conditioning on rhythm patterns. Finally, we evaluate the proposed model using quantitative analysis of rhythm and the subjective listening study. Results show that the music generated by our model tends to have salient repetition structures, rich motives, and stable rhythm patterns. The ability to generate longer and more structural phrases from disentangled representations combined with semantic scenario specification conditions shows a broad application of our model.
\end{abstract}

\IEEEpeerreviewmaketitle

\section{Introduction}
With recent breakthroughs in artificial neural networks, deep generative models have become the leading techniques for automated music generation \cite{Briot}. Many systems have generated more convincing results than traditional rule-based methods \cite{Loy}. Despite the promising progress, generating naturally structured music remains a challenging problem. Music structures of most compositions are complicated and involve multiple levels of groupings like repetitions, variations, and rhythmic conversions \cite{GTTM}. It is worth noting that most successful existing cases of generating human-like compositions are limited to EMI \cite{DavidCope}, Bach's style mimicking models \cite{Deepbach, Bachbot}, Memex \cite{memex}, and the Music Transformer \cite{AnnaHuang}. EMI managed to solve both structure and content problems in generating several classical compositions. However, it requires a very detailed analysis of existing compositions, which is time-consuming and at a disadvantage in generative speed. Bach's style models can create beautiful polyphonic pieces in a second. But the structure of Bach's chorales is easy to perceive, which makes these models not be generalized to other composers' works. Memex is based on Factor Oracle that requires manual feature selection and symbolization. In that it lacks the feature learning and disentanglement of music. The Music Transformer performs well in large varieties of music styles, but the "midi message encoding" it proposed still does not resolve the issues of structure. The motives of generation are complex enough while in most cases, its rhythm and repetition fall into final blindness. 

In this paper\footnote{\href{https://github.com/RetroCirce/Auto-mask-Music-Generative-Model-via-EC2-VAE-Disentanglement}{The model is available in https://github.com/RetroCirce/Auto-mask-Music-Generative-Model-via-EC2-VAE-Disentanglement}}, we seek to address the structural missing by utilizing explicit condition embedding \cite{MMRP} and disentangled latent vectors of note sequences into one whole generative system. We specifically design a continuous melody writing scenario that machine should continue writing the future melodic pieces based on the given music information at the beginning. This generative situation has already been applied to many cases in other areas like image generation \cite{pixelRNN}, audio synthesis \cite{Wavenet} and dialog system \cite{DialogPos}. For music, continuous melody writing can be introduced to many use cases:  

\begin{enumerate}[itemsep = 0pt,parsep = 0pt,leftmargin = 20pt,topsep = 0pt,partopsep = 0pt,]
    \item inspiration aid that people can explore the possibilities of future melodies with the existing information they produce and the certain future constraints they set (structural conditions); 
    \item the interactive performance that people can transfer their perception of a structural pattern into the generation of music (e.g. tapping the rhythm of drums and then transferring it into the rhythm of the piano).
\end{enumerate}

The novel approach we propose is the combination among Conditional Variational Auto-encoder for the representation, continuous melody writing in the generative design and the usage of disentangled technique in creating conditions. We implement EC$^2$-VAE \cite{nime, vaeismir} into the pre-processing of our music representation. Specially, we leverage the encoder of VAE to compress a note sequence of the 8-beat period into one latent vector. In that, the following generative model can predict melodies in terms of short-term music sentences. Furthermore, the latent vectors via EC$^2$-VAE have separated contents in different ranges of dimensions via disentanglement technique, namely the pitch dimensions and rhythm dimensions. 

Several previous works have already utilized the latent vector of VAE into the music composition process via interpolation \cite{MusicVae} and in-painting \cite{vaeinpainting}. But this is the first time we incorporate the disentanglement technique into the condition of music generation in each step. This bridges the hidden structural representation and the actual structural performance (i.e. in our case, the rhythm pattern). Also, the first stage of our model receives the series of 8-beat compressed vectors and creates memory connections among longer music phrases, instead of note by note as many existing models produce. This will promise more convincing and larger repetitions throughout the whole generation. These two are the main contributions of our model.

\begin{figure}
 \centering
 \includegraphics[width=\columnwidth]{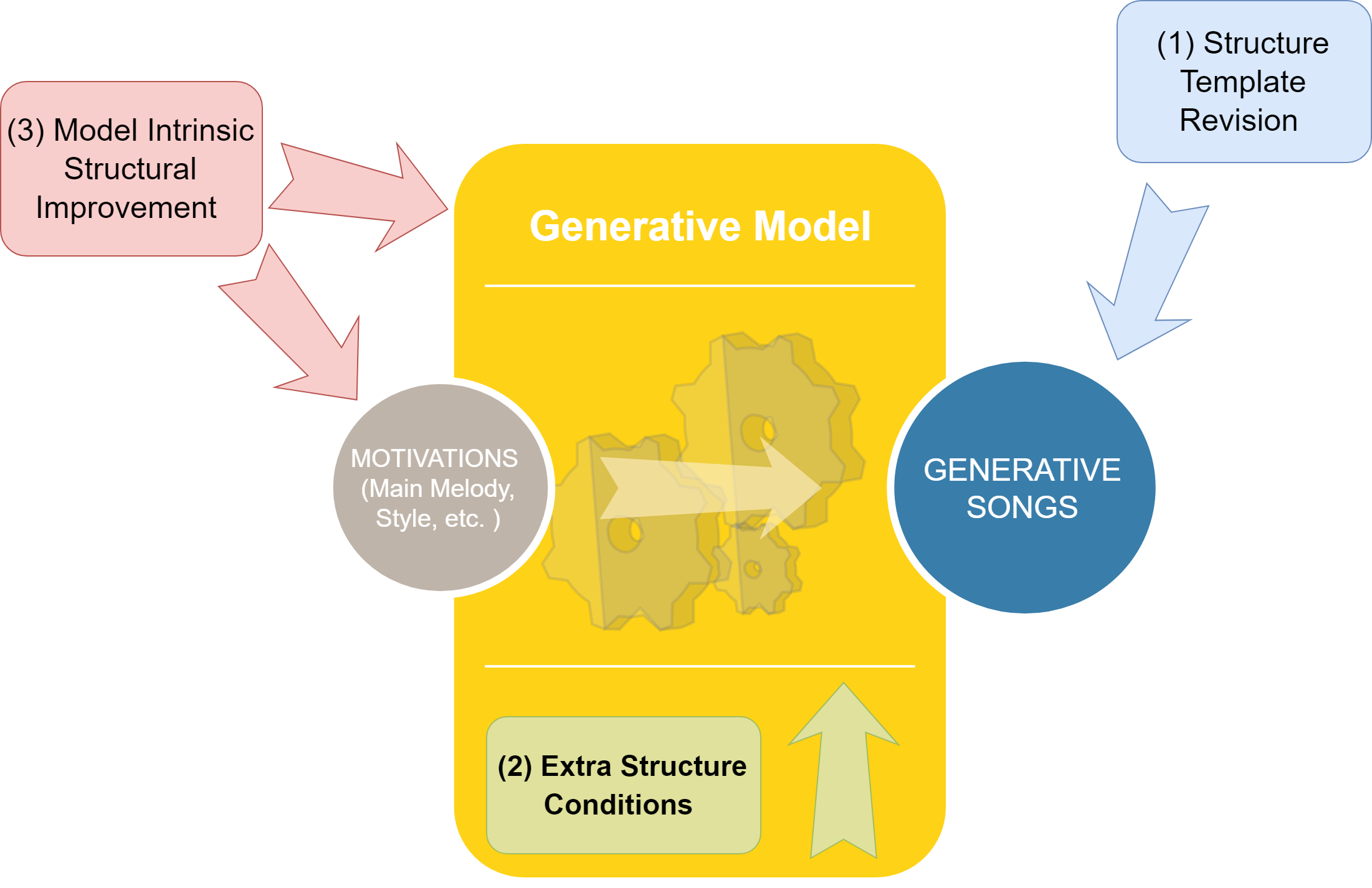}
 \caption{Three approaches to solving the structural problem.}
 \label{fig:structure method}
 \vspace{-0.5cm}
\end{figure}

\section{Related Work}\label{sec:page_size}

\subsection{Structural Construction on Generation}

As shown in Figure 1, three approaches can be used to solve the structural problem from a systematic perspective:

\begin{enumerate}
    \item post-processing the output by structural templates \cite{StructureNet,ImposingStructure}; 
    \item incorporating the inputs with extra structural conditions \cite{MMRP};
    \item making the model intrinsically more structural \cite{JapanTree}.
\end{enumerate}

The first approach is conceptually straightforward but often needs to generate a large number of candidates to be filtered by the templates. Moreover, the templates sometimes distort the musicality of the outputs when forcing them to match specific structures. The second approach is a smart shortcut to impose long-term structures (e.g. a lead sheet for jazz improvisation) since there is a little disagreement on them and even human composers create music “conditioned” on the given forms. The last approach is the most difficult, which requires the model to have the ability to represent music structures, and so far we see very few deep generative models capable of learning structural representations. 

In this paper, we combine the second and third approaches above. To be specific, the model contains two conditions in the second stage. The first is the setting of \textbf{the symbolic scenario}. This concept is introduced in \cite{impro-nika-improtek} by using structural representations to guide the music generative process. Any meta-data or label can be denoted as the structure representation. In this paper we add the manual labeling vectors of chord functions. This serves as the external condition to indicate the role of each generative phrase within the timeline. The second is the rhythm pattern hidden vector extracted by \textbf{the disentanglement technique}. This does not explicitly show how the rhythm will be in the current step. But the generation combined with this condition will share the intrinsic rhythm pattern with the song we provide as a template. The key to this combination is that we mimic human composers to constrain the overall progression of music with simple music theory. Meanwhile, we incorporate abstract music patterns by transferring from other existing music, as we called "imitation".

\begin{figure}
 \centering
 \includegraphics[width=6.5cm]{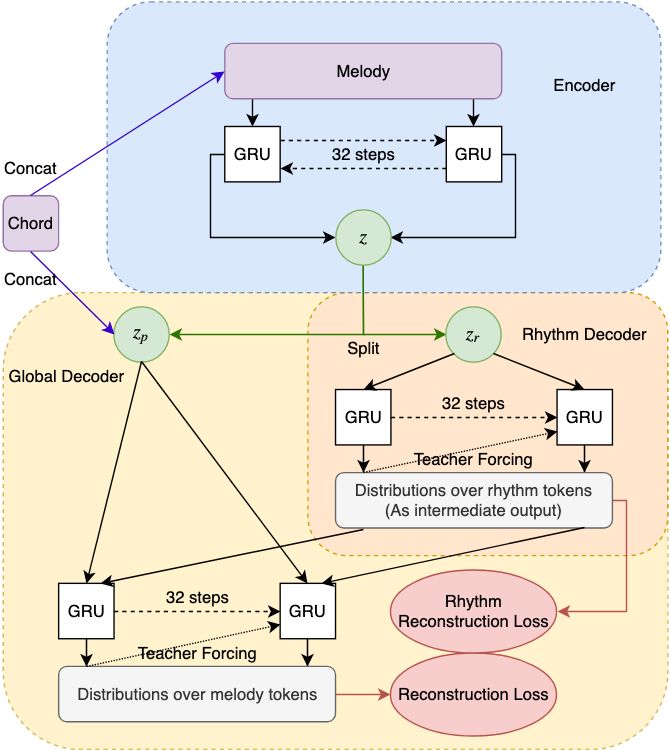}
 \caption{Reproduced from \cite{vaeismir}. The structure of EC$^2$-VAE for hierarchical music representation.}
 \label{fig:music-EC$^2$-VAE}
 \vspace{-0.6cm}
\end{figure}

\subsection{Short-Term Representation via VAE}

Variational Auto-encoder (VAE) has become a popular method of compressing high dimensional data into lower but meaningful vectors based on the Gaussian mixture model. In the aspect of music, several works have already introduced and implemented VAE into the representation of long musical sequences \cite{nime, MusicVae}. The encoder of VAE can output the latent vector of the musical sequence and the decoder can reconstruct it to the original sequence. 

We implement EC$^2$-VAE \cite{vaeismir}, the latest VAE for musical usage, in the encoding and decoding of music data. Figure 2 shows the structure of EC$^2$-VAE. Briefly speaking, EC$^2$-VAE can receive a series of notes as melody data, conditioned with optional chord data. Then its encoder will not only compress the melody data into lower-dimensional latent vectors but also divide the latent vector into different meanings. The method of its disentanglement is hierarchically calculating various loss functions in different ranges of vector's dimensions. The output of the encoder, defined as $z$, is a 256-d vector composed of two independent vectors $\{z_{p},z_{r}\}$. Then again with the optional chord data, its decoder can reconstruct the melody data. 

The latent vector $z$ reveals a representation of notes sequence. Moreover, $z_{p}$ independently denotes the pitch contour of melody and $z_{r}$ denotes the rhythm pattern. This successful separation provides a more hierarchical output object in our generative model.

\section{Methodology}\label{sec:page_size}

\subsection{Problem Definition}
\begin{figure}
 \centering
 \includegraphics[width=\columnwidth]{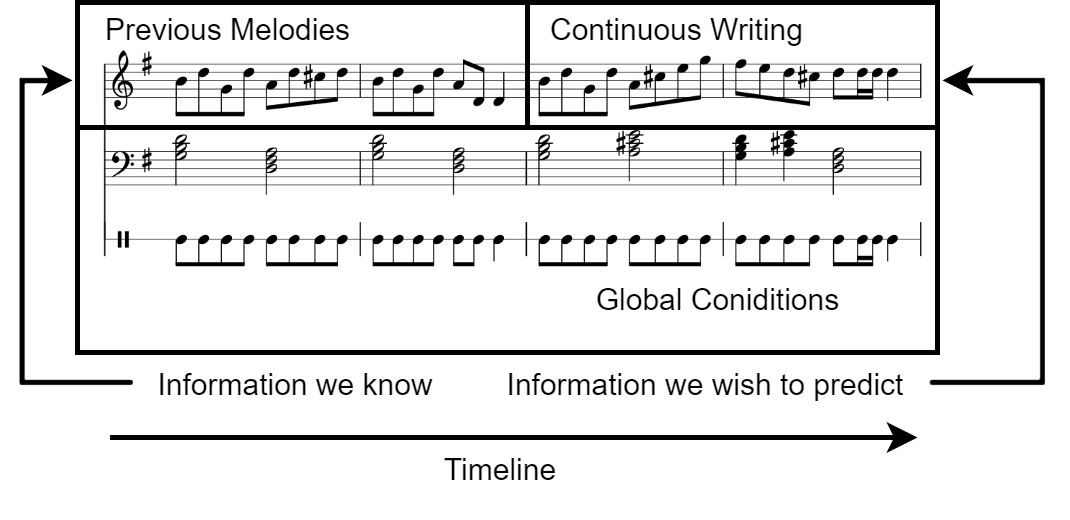}
 \caption{Continuous melody generation scenario. The future melodies are the information we wish to predict. The conditions like the chord progression, the chord function, and the rhythm section are set throughout the timeline for a better generation.}
 \label{fig:music-EC$^2$-VAE}
\end{figure}

Similarly to \cite{MMRP}, we put the melody generation task in the continuous writing scenario where previous music information is known and there is at least one global condition throughout all the timeline. And the model is required to generate the following melodies given both previous and global conditions. 

The conditional probability distribution of this task is defined as:
\begin{equation}
\begin{aligned}
    p(M_{t+1:T}\,|\,C_{1:T}) = \prod_{i=t+1}^{T}p(m_i | m_1,...,m_{i-1}, c_1, ..., c_T)\\
\end{aligned}
\end{equation}
where $M_{1:T} = \{m_1, m_2, ..., m_T\}$ denotes the melody in each time step. $C_{1:T} = \{c_1, ..., c_T\}$ denotes the condition in each time step. $t$ is the length of previous known information. And $T$ is the total length of whole generation including known information. 

As shown in Figure 3, the information we know and wish to predict is illustrated in detail. The condition listed above can be everything like chord progression, beat location or even none. The choice of conditions will directly lead to the final performance and application of generation.  

\subsection{Data Representation and EC$^2$-VAE Utilization}

As shown in Figure 4, we introduce \textit{holding}, \textit{rest} and \textit{pitch} states from \cite{MMRP,nime,MusicVae} to represent the original symbolic melody as a 130-dimension one-hot melodic vector $m_i$ in each time-step $i$. The first 128 dimensions are regarded as pitch value (0 - 127). The \textit{rest state} implies that the empty note in current timestep. And \textit{the holding state} represents the duration of the previous pitch. Then we implement a 12-dimension chord vector $chord_i$ as chromatic feature for the chord. 

For the condition preparation in the symbolic scenario, we incorporate chord function labels in the representation of music data. The chord function is used to denote the relationship between a chord to the tonic center of a song. The chord function plays a great indicative role in the transition and coherence of musical phrases. The tonic center of a song depends on the scale (Major or Minor). Any chord function label of a song is in line with it. Here we define four labels: T (Tonic), D (Dominant), S (Subdominant) and O (Other). According to the chord function theory \cite{HugoR}, we assign the chord function label in Figure 5. Note that we simplify the classification to four slots (T, D, S, O). Some degree chords might not be classified very correctly (e.g Major VI chord is defined as \textit{Tonic Parallel} on major scale but we assign it as Tonic). The labels we use do not greatly deviate from the original function of chords. 
\begin{figure}
 \centering
 \includegraphics[width=7.5cm]{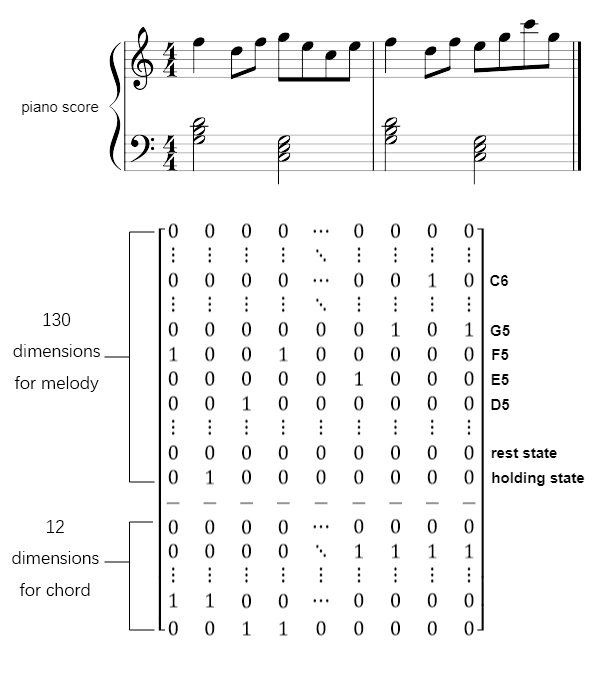}
 \caption{Data representation of pitches and chords, namely 130-dimension for pitches and 12-dimension for chords.}
 \label{fig:dt-mc}
\end{figure}
\begin{figure}
 \centering
 \includegraphics[width=\columnwidth]{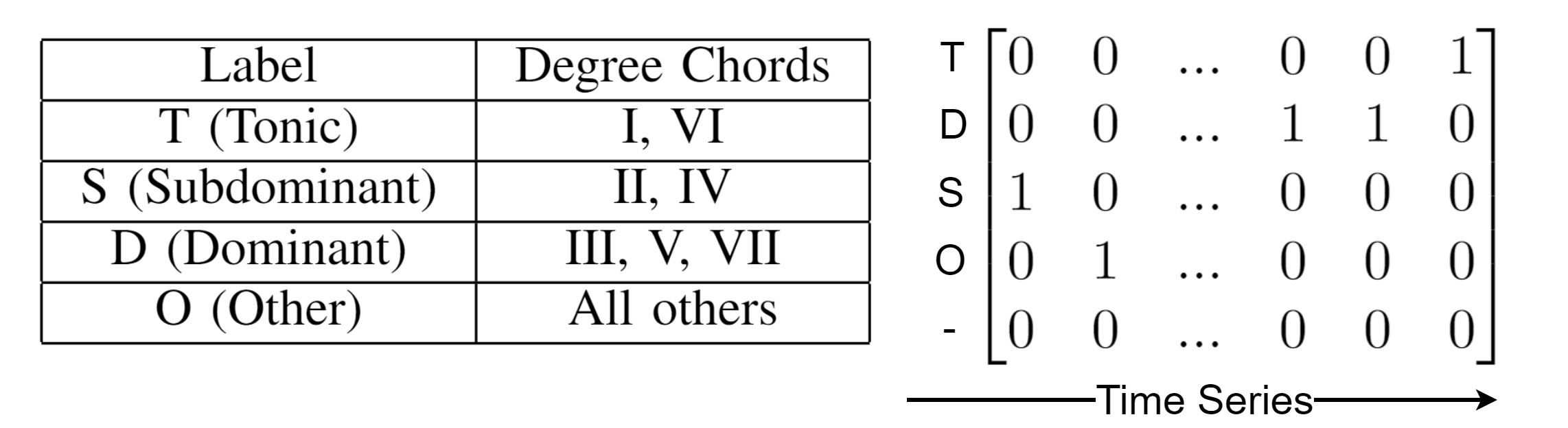}
 \caption{Chord function labels of different degree chords and the explicit structure of labels.}
 \label{fig:chord-function-label}
\end{figure}

We provide a C-major examples for better understanding:
\begin{itemize}[itemsep = 0pt,parsep = 0pt,leftmargin = 10pt,topsep = -1pt,partopsep = 0pt]
    \item When the scale is C-major:
        \begin{itemize}[itemsep = 0pt,parsep = 0pt,leftmargin = 20pt,topsep = 0pt,partopsep = 0pt]
            \item C, Am are Tonic (with their corresponding 7th and 9th chords like Cmaj7, Am7)
            \item F, Dm are Subdominant (with 7th and 9th chords)
            \item G, Bdim, Em are Dominant (with 7th and 9th chords)
            \item All others are Other.
        \end{itemize}
\end{itemize}

As shown in Figure 5, we use five-dimension one-hot vectors to represent the chord function in each timestep of the music data. The first four dimensions are for T, D, S and O. The last one dimension is just for padding (no use).

\subsection{Model Architecture}

In this section, we will introduce the architecture of the whole model, including the representation process (the first stage) and the conditional generative process (the second stage). 

\subsubsection{Representation Encoding via EC$^2$-VAE}
Rather than generating the melody note by note directly by $\{m_i\}$, we feed the sequence data $\{(m_i, chord_i) | i = 1...n\}$ into EC$^2$-VAE's encoder by the length of $n$ timesteps. The output of the encoder, denoted as $z$ is the latent representation of a specific melody sequence with chord progression.
Furthermore, the latent representation $z$ is a 256-dimension vector composed by pitch contour vector $z_{p}$ and rhythmic pattern vector $z_{r}$. 

In that, on the generative process, we do not use the time-step-based representation $\{m_i\}$ for predicting melodies note by note. We use $z$ as the basic unit to predict melodies phrase by phrase, which is more close to what musicians compose in the creative process.

\subsubsection{Conditional Prediction via Bi-LSTM}
We introduce 8-layer Bi-LSTM with short-cut \cite{ResNet} in the melody generation task. The reason for our choice to the number 8 of layers is because we wish to make the model deeper and capable of handling generation, meanwhile, make it efficient enough to reduce extra cost. After several attempts on training with different numbers of layers, 7-8 layers stand out to be the best proper one for our task and dataset. With the advantage of VAE and conditional embedding. The prediction task can be revised and specified as:
\begin{equation}
\begin{aligned}
    &p(Z^p_{t+1:T}\,|\,(CF_{1:T},Z^r_{1:T})) \\
    &=\prod_{i=t+1}^{T}p(z^p_i| z^p_1,...,z^p_{i-1},z^r_1,...,z^r_T,cf_1,...,cf_T)\\\
\end{aligned}
\end{equation}
where $Z^p_{1:T} = \{z^p_1, z^p_2, ..., z^p_T\}$ denotes the pitch contour of each melodic sequence. $CF_{1:T} = \{cf_1, ..., cf_T\}$ denotes the chord function condition. And $Z^r_{1:T} = \{z^r_1, z^r_2, ..., z^r_T\}$  the rhythm pattern of each melodic sequence as another condition throughout all the timeline. One thing to be noted is that in this case, $T$ and $t$ no longer denote the length of time steps, but the length of ordered groups in time steps. Each group contain $n$ time steps melodies as we defined before.  

Note that when the sequence is restricted with extra conditions (i.e. chord function, rhythm pattern, etc.), Bi-LSTM can help the model utilize conditions from start to end. The generation in the current time group can not only concern about the previous information but also think of the future conditions \cite{MMRP}. In this case, the large part of the rhythm pattern as one of the music structure is therefore preserved and contributed to the prediction of future pitch contents.

\begin{figure}
 \centering
 \includegraphics[width=\columnwidth]{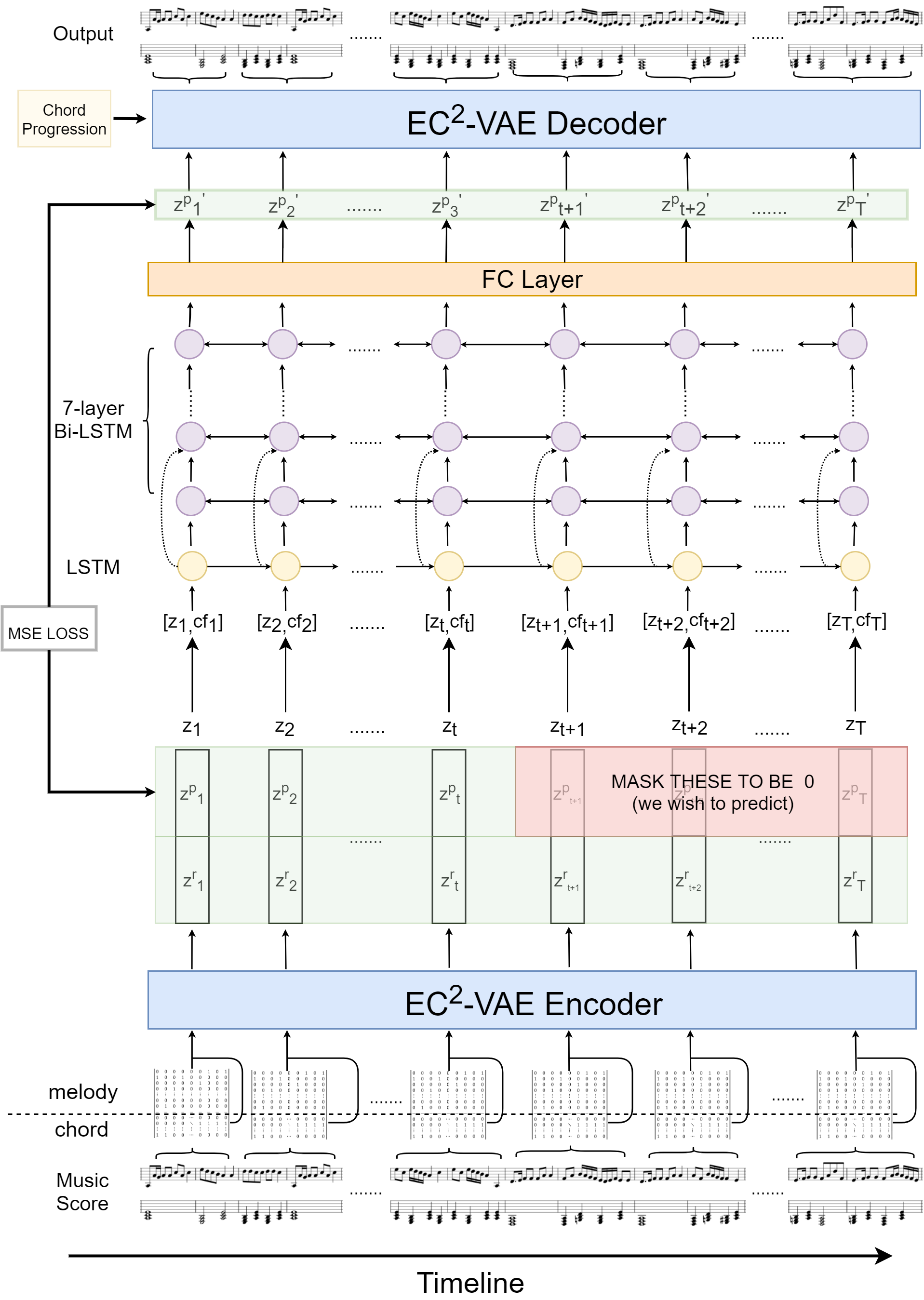}
 \caption{The whole model architecture of the proposed model. The first stage (Both on the bottom and top of the figure): EC$^2$-VAE representation encoding. The second stage (in the middle of the figure): conditional LSTM generative model.}
 \label{fig:chord-function-label}
\end{figure}
\begin{figure*}
 \centering
 \includegraphics[width=17cm]{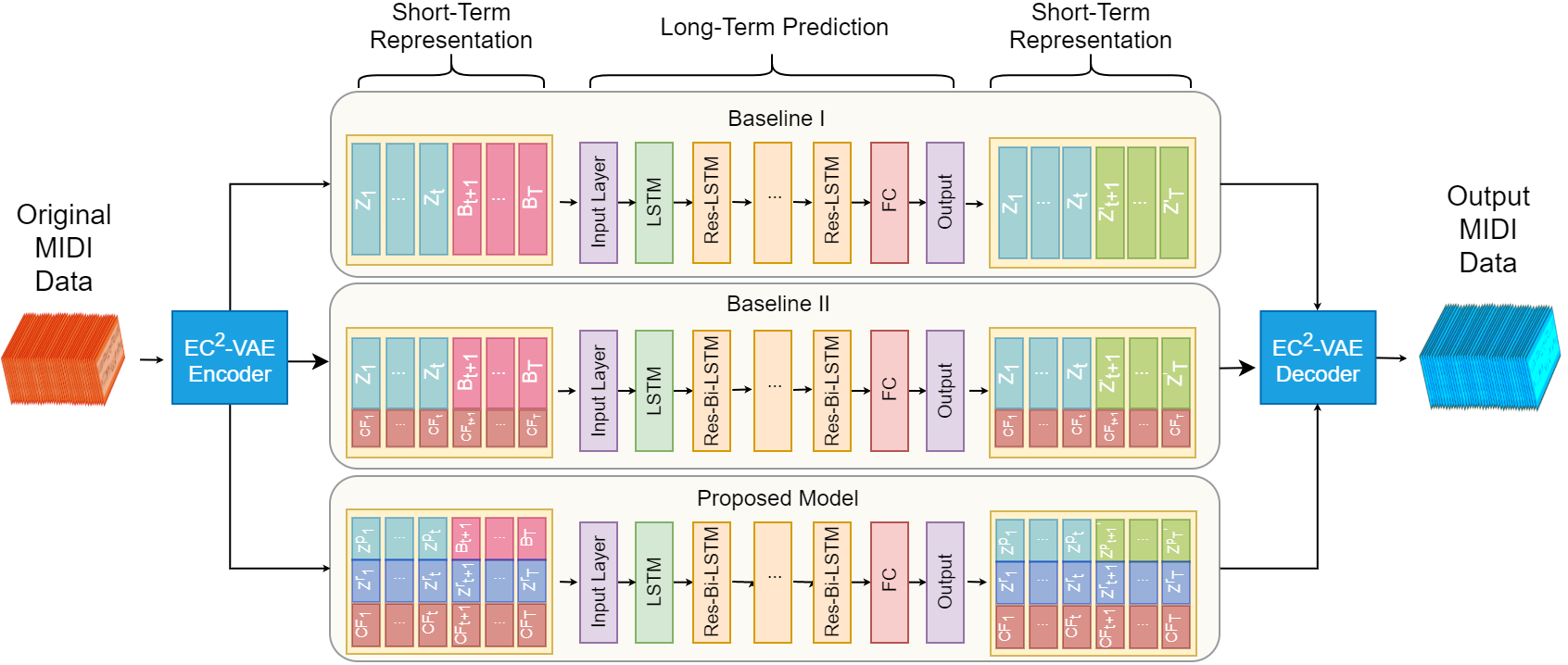}
 \caption{Architectures of three models. In each model box, we can see the $blank_i$ or $B_i$ (marked by light red) in the left and $z'_i$ or $z'^{p}_{i}$ (marked by light green) in the right. The "light red" is what we leave empty for prediction and the "light green" is the generative result.}
 \label{fig:model arch}
\end{figure*}
\section{Experiments}\label{sec:typeset_text}

\subsection{Dataset, Baseline and Training Details}\label{subsec:body}

We choose Nottingham Database \cite{Nottingham}, a collection of 1200 British and American folk songs, as the training source of our models. We implement the data augmentation by switching all songs to 12 tonalities. And we normalize all songs to 120 bpm (beats per minute) and set the $timestep = 0.125 \,secs$ ($\frac{1}{4}$ beat in $\frac{4}{4}$ time signature) in order to get better alignments. 

We feed music data (melodies and chord progressions) into EC$^2$-VAE. We set $n = 32$ (the length of the sequence), so that every 8-beat (0.125 x 32 = 4.0 secs) melodic piece will be decoded as one $z$. We use Cross-Entropy as the loss function between each original and generative piece:

\begin{equation}
    H(M_{1:n}, M'_{1:n}) = -\sum_{i=1}^{n} m'_{i}\,log\,m_{i}
\end{equation}
where $M_{1:n}$ is the original melody sequence $\{m_1,m_2,...,m_n\ | n = 32\}$ and $M'_{1:n}$ is the reconstructed melody $\{m_1',m_2',...,m_n'\ |\,n = 32\}$ decoded from $z$.

Then we utilize these hierarchical representations in the generative model as we describe in Figure 6. We set the total length $T = 10$ and the known length $t = 5$ in the generative model. It means that we use the previous five latent vectors (4.0 x 5 = 20.0 secs) to predict the succeeding five latent vectors (20.0 secs). And the global conditions (chord function labels and $z_{\text{rhythm}}$) are considered throughout the whole 40.0 secs sequence. 

As shown in Figure 7, two baseline models are incorporated to make comparisons. The first baseline model on the top is the LSTM model without any condition in the future. The second baseline model in the middle constrains the future prediction only by the explicit chord function label. The model on the bottom is our design, which combines the explicit chord function label and the rhythm pattern vector in the latent space. You can easily understand the process by recognizing that the red boxes marked as $B$ are the elements we wish to predict. As you can see, they turn to green in the right after finishing the prediction, or continuous writing. Other boxes in colors are different conditions we set in different models ($z,z_r,z_p$ for the latent vector, $cf$ for the chord function label).

We pick all available sequences in the train data for three models. And we select sequences in the test data as generative beginnings to see if models can develop them to good results. All details about the experiment from representation to generation are shown in Table 1.

\begin{table}  
\newcommand{\tabincell}[2]{\begin{tabular}{@{}#1@{}}#2\end{tabular}}
\centering
\begin{tabular}{cc}  
\hline  
\hline  
EC$^2$-VAE CE Loss & 0.0050  \\  
\hline
\hline
Model No. & MSE Loss  \\  
\hline
Baseline I: EC$^2$-VAE + LSTM  & 0.0016 \\  \hline
\tabincell{c}{Baseline II: EC$^2$-VAE + Bi-LSTM \\ + Chord Function condition} & 0.0009  \\\hline
\tabincell{c}{Proposed Model: EC$^2$-VAE + Bi-LSTM \\ + Chord Function \& $z_{\text{rhythm}}$ conditions} & 0.0005 \\
\hline  
\end{tabular}  
\caption{Model training details including EC$^2$-VAE and generative models} 
\end{table}  

\subsection{Subjective Listening Study}
From the perspective of music generation, we expect models to create samples that are convincing to human beings. We conduct a multi-view survey for subjects to judge generations from different models based on different criteria. Throughout all the survey we compare three models: 1) LSTM model with any condition; 2) LSTM model with chord function labels as an explicit condition; 3) LSTM model with chord function labels and latent rhythm pattern vector $z_r$ as conditions (our proposed model).

Subjects are required to listen to one original song first by memorizing motives, repetitions, rhythm patterns, and other structural or non-structural content. Then they will listen to the generations of three models in a random order. The only same part of three generations is the first 20-second beginning (five $z$ we describe as "first previous music information"), which is copied from the original song. Then different developments appear as different models infer. Subjects rate of each song based on three criteria:
\begin{enumerate}
    \item Interactivity: Do the chords and melodies interact with each other well?
    \item Complexity: Are the notes complex enough to express the theme?
    \item Structure: Can you notice some repetitions, rhythm patterns, and variations that \textbf{share great consistency with the reference song}?
\end{enumerate}
And they still need to rate the overall musicality of generations.

\begin{figure}
 \centering
 \includegraphics[width=\columnwidth]{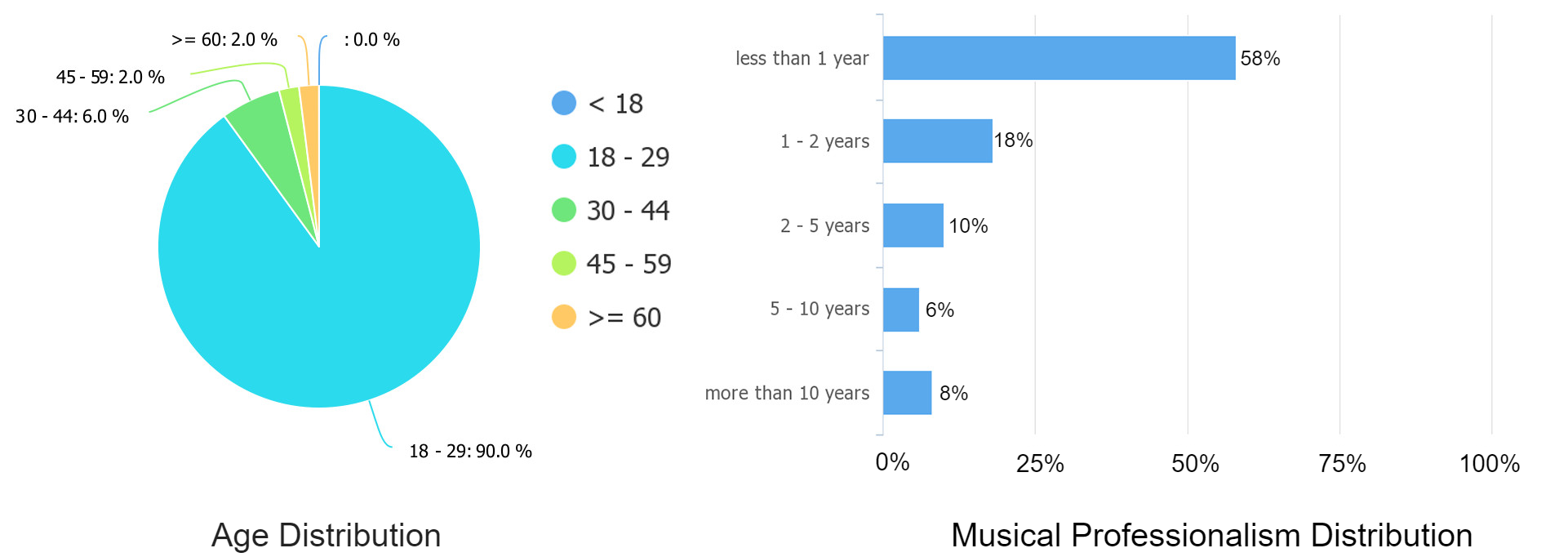}
 \caption{The age and the musical professionalism distributions of the subject listening survey. Reproduced from the Tencent Survey.}
 \label{fig:chord-function-label}
\end{figure}

\begin{figure}
 \centering
 \includegraphics[width=\columnwidth]{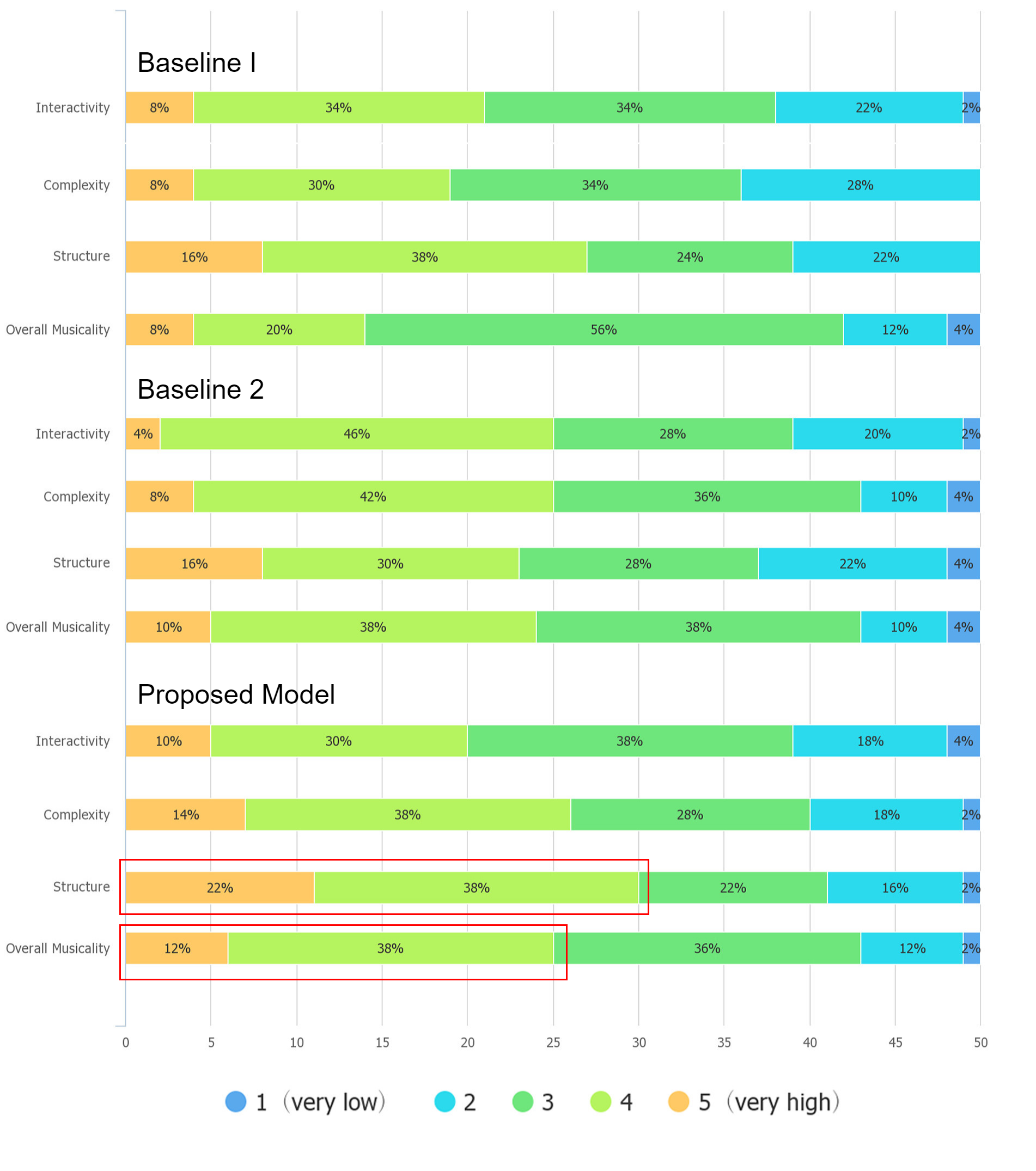}
 \caption{Rating results from three criteria and overall musicality of three models. We can find by the red boxes that our proposed model achieves almost all the highest number of 5-star and 4-star, especially in Structure and Overall Musicality. Reproduced from the Tencent Survey.}
 \label{fig:chord-function-label}
\end{figure}
We collect 50 subjects, 200 ratings in total. The gender, age, and musical profession distributions are shown in Figure 8.

Then, as shown in Figure 9, the rating result contains 4 views. In our case, we mainly focus on Structure criteria and Overall Musicality. We can find by the red boxes that our proposed model achieves almost all the highest number of 5-star and 4-star, especially in Structure and Overall Musicality. Combined with Table 2, which displays the average ratings of three models, universal results show the promising generations of our model in terms of rhythm and musicality to human beings.

\begin{table}  
\newcommand{\tabincell}[2]{\begin{tabular}{@{}#1@{}}#2\end{tabular}}
\centering
\begin{tabular}{cccc}  
\hline  
\hline  
Model No. & Baseline I & Baseline II & Proposed Model \\  
\hline
\hline
Interactivity & 3.24 & 3.3 & 3.24 \\ \hline
Complexity & 3.18 & 3.4 & \textbf{3.44} \\ \hline  
Structure & 3.48 & 3.32 & \textbf{3.62} \\ \hline  
Overall Musicality  & 3.16 & 3.4 & \textbf{3.46} \\ \hline  
\end{tabular}  
\caption{The average ratings of three models in terms of all criteria.} 
\end{table}  

\subsection{Quantitative Analysis for Rhythm}

\begin{figure}
 \centering
 \includegraphics[width= 8cm]{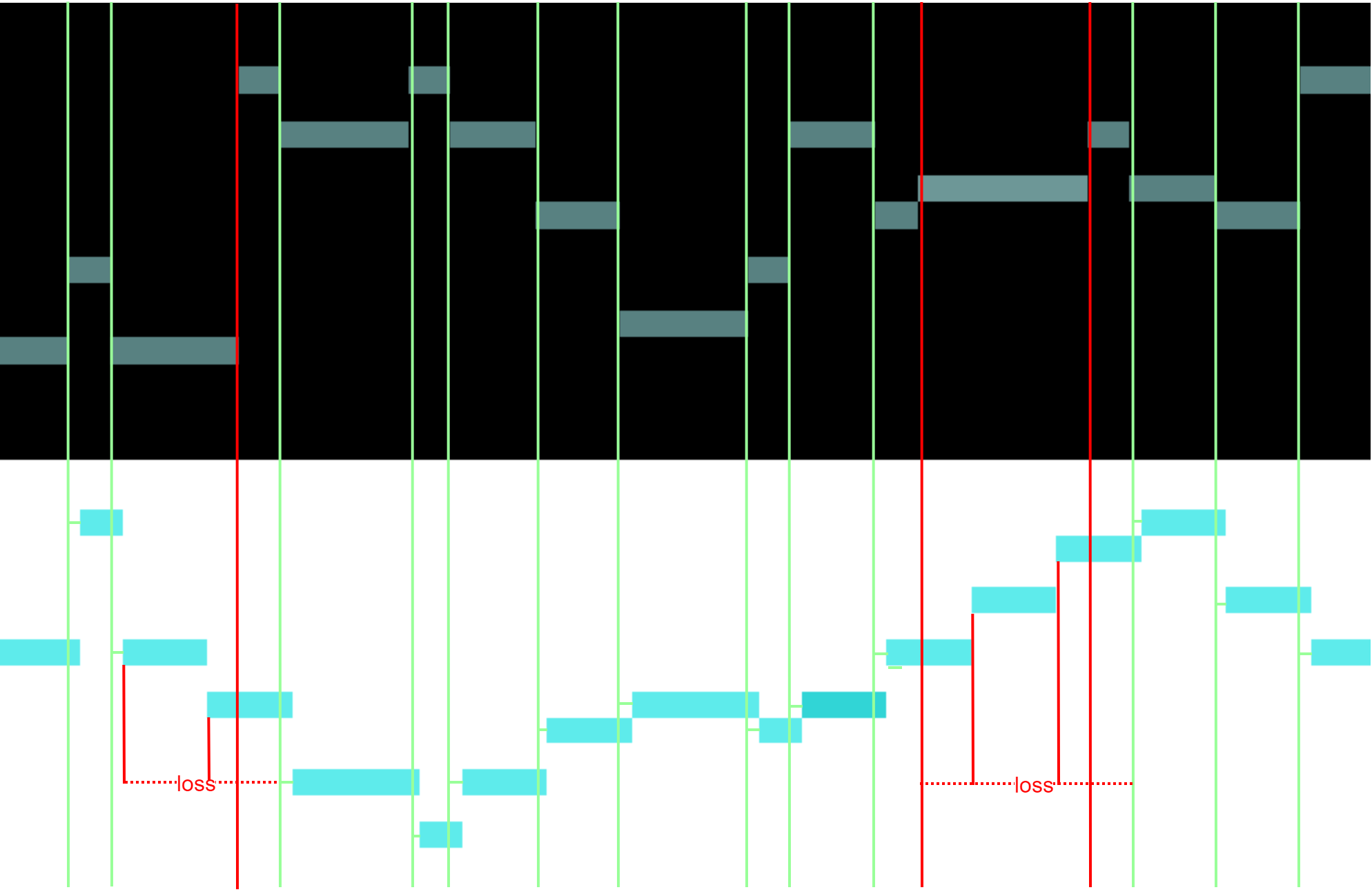}
 \caption{The illustration of how rhythm accuracy is calculated. Green lines denote the number of onset hits from each note of the reference song within tolerance. Red lines denote the failures of hits. Red segments denote the places we should calculate the loss of onset.}
 \label{fig:chord-function-label}
\end{figure}
Even though the subjective listening study highly reflective the average quality of the model's generation overall, a more scientific and engineering-based method should be introduced to prove the rhythmic preservation of our model. In this section, we propose a duration-ratio-based algorithm to calculate the onset accuracy in relation to the rhythm pattern. 

As shown in Figure 10, we calculate the onset accuracy of one generation given the reference song (the original data). Green lines denote the number of onset hits from each note of the reference song within tolerance. Red lines denote the failures of hits. Red segments denote the places we should calculate the loss of onset. We define the formula as:
\begin{equation}
\begin{aligned}
    &Rhythm\_Accuracy(refer, generation) = \\
    &\frac{1}{n}\sum_{i=2}^{n}(\delta (-\epsilon < r_i - g_{h(r_i)}) < \epsilon) \times (1 - \frac{\sum_{j = h(r_{i-1})+1}^{h(r_i)-1}d_g(j)}{\sum_{j = h(r_{i-1})}^{h(r_i)}d_g(j)}))
\end{aligned}
\end{equation}
where $r_i$ denotes the onset of $i^{th}$ note in the reference song, $h(x)$ denotes the index of the note in the generation which has the least distance to the time $x$, $g_i$ denotes the onset of $i^{th}$ note in the generation, $d_g(i)$ denotes the duration of $i^{th}$ note in the generation, and $\delta(cond)$ denotes the sign function (1 if cond = True else 0), and the $\epsilon$ is the onset tolerance, which is set to 0.1-sec in our experiment.

Generally speaking, the accuracy function will calculate each contribution in the note of the reference song in terms of the ratio of duration if it successfully hits the generation. If the onsets of the generation are similar to those of the reference, the accuracy will go up. Otherwise, each hit loss and the "extra" note between hits will be counted into the loss of the whole accuracy.

\begin{table}  
\newcommand{\tabincell}[2]{\begin{tabular}{@{}#1@{}}#2\end{tabular}}
\centering
\begin{tabular}{cccc}  
\hline  
\hline  
Model No. & Baseline I & Baseline II & Proposed Model \\  
\hline
\hline
Rhythm Acc & 0.809 & 0.778 & \textbf{0.981} \\ \hline
\end{tabular}  
\caption{The rhythm accuracy calculated in the test data of three models. The results are highly consistent with the Structure criteria in the subject listening study.} 
\end{table} 

We calculate each generation from three models (two baselines and our proposed model) with original data as reference songs. Table 3 shows the average results we get about the rhythm accuracy. We can easily find that there is a clear gap between the rhythm accuracy of different models. Our model almost completely reproduces the rhythm of the original song due to the rhythm pattern we transfer in the latent vector as a condition. Therefore the generation of our proposed model can have greatly promised rhythm patterns consistent with the original song.

\begin{figure}
 \includegraphics[width=\columnwidth]{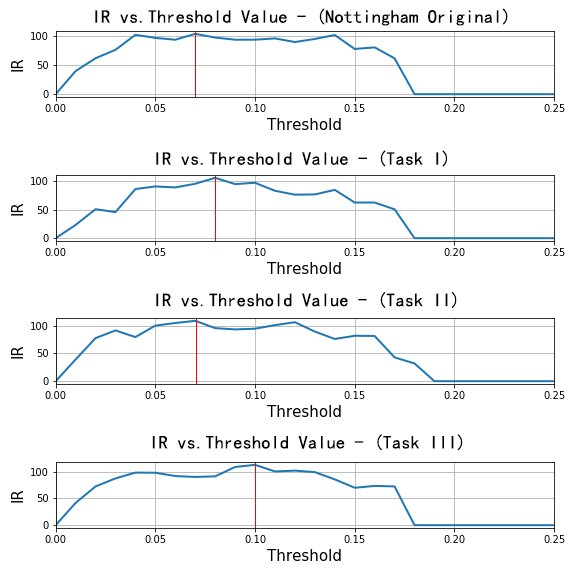}
 \caption{IR vs. Threshold $\theta$ value (VMO).}
 \label{fig:IR threshold}
\end{figure}
\begin{figure}[!h]
 \includegraphics[width=\columnwidth]{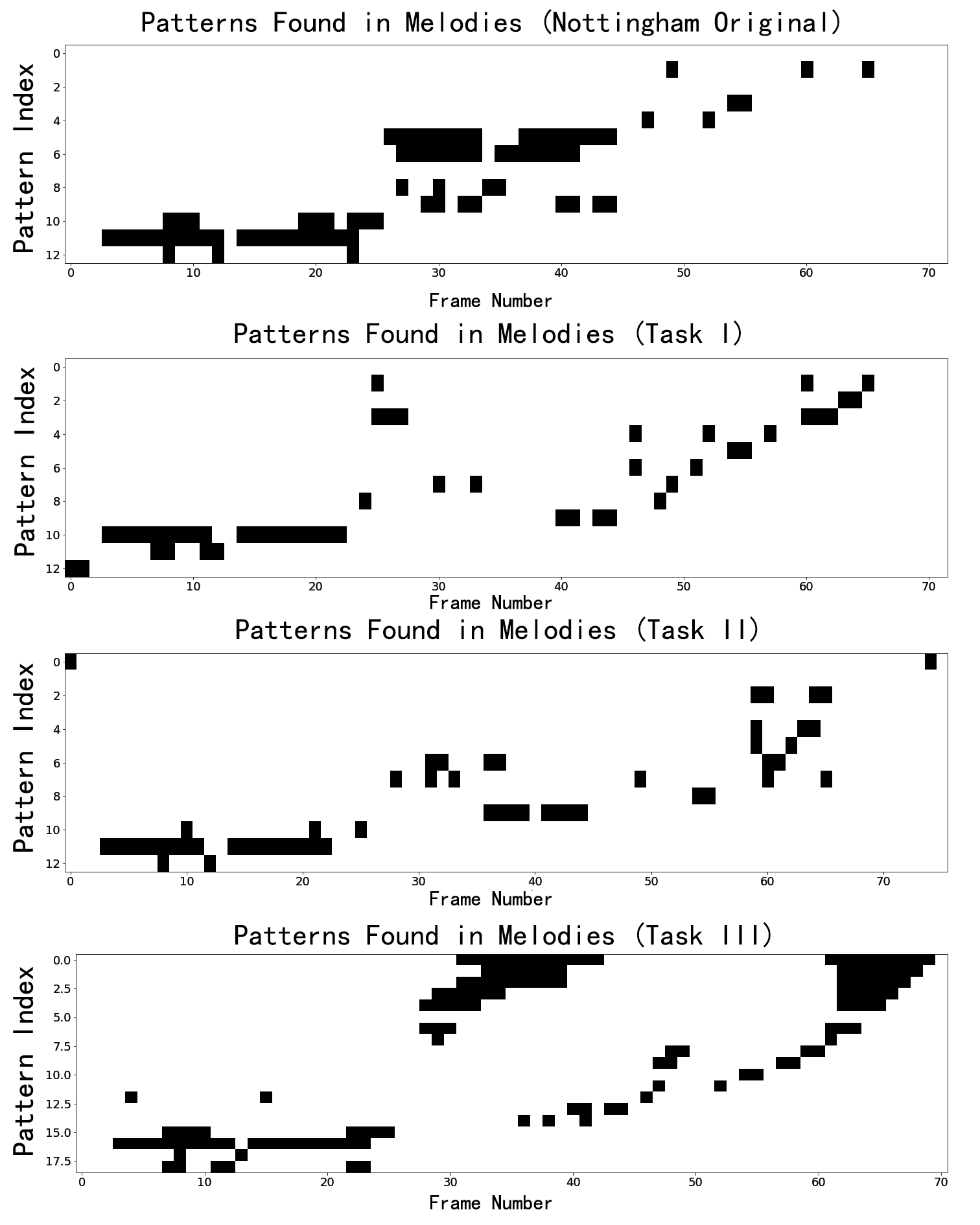}
 \caption{Pattern Discovery with VMO.}
 \label{fig:Pattern VMO}
\end{figure}

\subsection{Repetition and Motif Discovery via VMO}

Variable Markov Oracle (VMO)\cite{VMO} is a suffix-based structure that can detect the repeated pieces of the given time series. In the analysis of music sequences, VMO can capture dynamic motives and repetitive structures. The composing ability of models in pattern creation and complexity can be extracted from it.

\begin{figure*}
\centering
 \includegraphics[width=18cm]{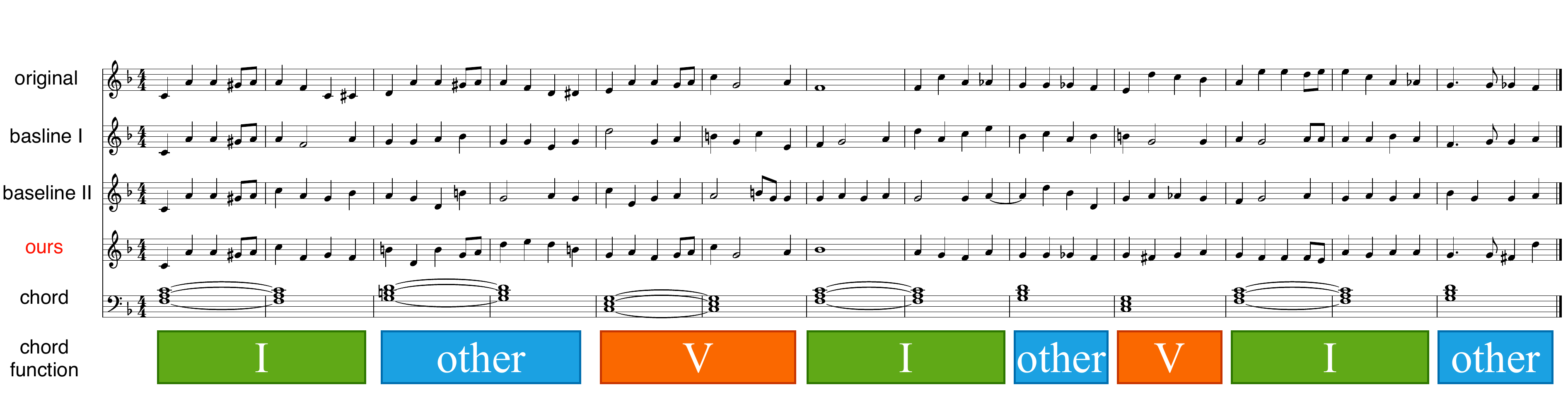}
 \caption{Music generation samples from three models, including baselines and our proposed model}
 \label{fig:Pattern VMO}
 \vspace{-0.5cm}
\end{figure*}

We implement Information Rate (IR) \cite{IR} in the choice of the comparison threshold. We pick the threshold to maximize the IR of each song. Figure 11 shows the visualization of the IR values versus different threshold $\theta$ of one song. From top to bottom we can see the original sample and three tasks' samples. Each red column marking indicates our choice of $\theta$. We can see all generations are little different from Nottingham Original data in the degree of changes of IR Curve. All model tasks we propose can generate melodies that are almost identical to the original song from the perspective of musical phrase length. In this sample, the maximum IR point of Task III (2.4.3) lies in the number which is a little larger than others'. This means that Task III has longer segments. Although Task III drops down its IR with a sharper line, the previous curve is smoother which indicates the longer blocks.

The analysis above can be further reflected in Figure 12, showing the VMO pattern discoveries of samples. The x-axis indicates the frame number within the time series. The y-axis indicates the pattern indexes we detect from each data. The musical complexity and repetitive structure can be reflected differently in the comparison of four graphs. We can recognize the repetitive motives and structural echoes in each model's sample. Our proposed model successfully takes on the task of memorizing and recalling the previous musical phrases. Besides that, it reveals a strong ability in composing music with larger complexity and longer developments.

\section{Discussion}

We utilize the hierarchical and disentangled representation into the generative model. And we set conditions based on both external explicit chord function labels and latent rhythm pattern vectors. The first merit of this realization lies in the ability of the model to generate long music more easily. As shown in Figure 13, generating about 12 bars of music pieces needs 2-3 generative iterations (20.0 secs per iteration). In the note-by-note scenario, this will cost more iterations. When it comes to the comparison among the three models, the generations reveal differences in music developments and rhythm patterns. As shown in Figure 13, the first three tracks are the generations by three tasks given the same previous information and global conditions. The bottom track is the chord progression. We can see Baseline I could only generate the melody with a stable rhythm and in a small range of notes. The range issue has been resolved in Baseline II as we incorporate the chord function condition, telling the model about where to development the motives. And in our proposed model, we preserve the rhythmic patterns. The generation tends to actively show the changes of rhythms. Besides that, the computational cost of our proposed model will be reduced because we only need to generate half dimensions of latent vectors. In that, this more lightweight prediction provides a potential application because we can combine the output with different function labels and rhythmic conditions transferred by other resources of music or human's performances, making the full use of the generative results.

\section{Conclusion}
In this paper, we contribute a hierarchical melody generation framework, which combines the power of short-term representation via EC$^2$-VAE and structural conditional prediction to generate more structural music. We first manage to use the disentanglement technique in the generative process. The key feature of the model is that it mimics human composers to take music as phrases and sentences, as opposed to most algorithms that generate one note, one beat, or even a shorter subdivision at a time. With the further help of representation disentanglement techniques, we show that this generation framework is flexible: the target can be either purely pitch contours or whole melodies conditioned on any structural input, such as chord progression and their functions. We see this study as an important step towards the semantic understanding of music content and structural music generation.

\section*{Acknowledgement}
We would like to thank Cygames for the partial support of this research.

\bibliographystyle{ieeetr}
\bibliography{ICSC-ke}

\end{document}